\documentclass[preprint,prx,amsmath,amssymb,longbibliography,superscriptaddress]{revtex4-1}

\pdfoutput=1

\usepackage{graphicx}
\usepackage{amsmath,latexsym,tabularx}
\usepackage{xcolor}
\usepackage{multirow}
\usepackage[
  colorlinks=true,
  urlcolor=blue,
  linkcolor=blue,
  citecolor=blue
]{hyperref}
\usepackage{multirow,booktabs}
\usepackage{float}

\newcommand{\affilLL}[0]{Lincoln Laboratory, Massachusetts Institute of Technology, Lexington, Massachusetts 02421, USA}
\newcommand{\affilMIT}{Massachusetts Institute of Technology, Cambridge, Massachusetts 02139, USA}

\newcommand{\unit}[1]{\, \mathrm{#1}}
\newcommand{\sr}{Sr$^{+}$}

\newcommand{\specificthanks}[1]{\@fnsymbol{#1}}

\begin{document}

\title{A Brillouin Laser Optical Atomic Clock}

\author{William Loh}
\thanks{These authors contributed equally to this work.}
\affiliation{\affilLL}

\author{Jules Stuart}
\thanks{These authors contributed equally to this work.}
\affiliation{\affilMIT}
\affiliation{\affilLL}

\author{David Reens}
\affiliation{\affilLL}

\author{Colin D. Bruzewicz}
\affiliation{\affilLL}

\author{Danielle Braje}
\affiliation{\affilLL}

\author{John Chiaverini}
\affiliation{\affilLL}

\author{Paul W. Juodawlkis}
\affiliation{\affilLL}

\author{Jeremy M. Sage}
\affiliation{\affilLL}
\affiliation{\affilMIT}

\author{Robert McConnell}
\affiliation{\affilLL}

\date{\today}

\begin{abstract}

Over the last decade, optical atomic clocks \cite{Hinkley2013, Bloom2014, Brewer2019, Koller2017, Godun2014, Huntemann2016} have surpassed their microwave counterparts and now offer the ability to measure time with an increase in precision of two orders of magnitude or more. This performance increase is compelling not only for enabling new science, such as geodetic measurements of the earth, searches for dark matter, and investigations into possible long-term variations of fundamental physics constants \cite{Maleki2005, Ludlow2015, McGrew2018} but also for revolutionizing existing technology, such as the global positioning system (GPS). A significant remaining challenge is to transition these optical clocks to non-laboratory environments, which requires the ruggedization and miniaturization of the atomic reference and clock laser along with their supporting lasers and electronics \cite{Leibrandt2011, Koller2017, Liu2018}. Here, using a compact stimulated Brillouin scattering (SBS) laser to interrogate a {$^{88}$\sr} ion, we demonstrate a promising component of a portable optical atomic clock architecture. In order to bring the stability of the SBS laser to a level suitable for clock operation, we utilize a self-referencing technique to compensate for temperature drift of the laser to within $170\unit{nK}$. Our SBS optical clock achieves a short-term stability of $3.9 \times 10^{-14}$ at 1 s---an order of magnitude improvement over state-of-the-art microwave clocks. Based on this technology, a future GPS employing portable SBS clocks offers the potential for distance measurements with a 100-fold increase in resolution.

\end{abstract}

\maketitle



The ability to precisely measure time with a portable system has long been instrumental to navigation. The necessity for accurate and portable timekeeping inspired the development of Harrison's marine chronometer nearly 300 years ago and continues to this day, reflected in modern societal reliance on the GPS. Recently, optical atomic clocks with performance far surpassing that of the best microwave clocks have significantly advanced the precision with which time---and, equivalently, distance---can be measured. However, portable implementations of these optical clocks will require substantial modifications to the existing clock architecture, including miniaturization of the master laser whose performance is central to the operation of the clock. A key challenge is to maintain the frequency stability of the clock laser while reducing its size. This stability is required for the laser to (1) remain within the vicinity of the atom's narrow-linewidth transition during the period of time before the clock feedback is engaged (${\sim}$minutes) and (2) remain locked to the atomic transition between feedback cycles (${\sim}$ms). These stringent requirements have thus far eliminated all but bulk-cavity-stabilized (BCS) lasers \cite{Young1999, Jiang2011, Kessler2012}, which exhibit linewidths of $<1$ Hz but otherwise are unwieldy and prone to vibration \cite{Leibrandt2011, Davila-Rodriguez2017, Koller2017, Didier2019}, as possible candidates for clock interrogation.

A promising portable alternative to these BCS lasers has recently emerged via generation of SBS light in an ultrahigh quality factor (Q) resonator \cite{Grudinin2009, Lee2012, Kabakova2013, Loh2015, Otterstrom2018, Gundavarapu2019}. The SBS nonlinearity plays a vital role here as the stimulated phonons provide an additional filtering mechanism for laser noise \cite{Debut2000}, which enables an SBS laser to surpass the ultimate level of linewidth achievable through the resonator Q alone. In comparison to the BCS laser, the SBS laser offers the advantages of reduced cavity volume, operation without vacuum, the ability to be rigidly mounted to any flat surface, and a potentially higher tolerance to vibration. Despite these properties, the application of the SBS laser to state-of-the-art atomic physics has yet to be demonstrated, primarily due to the laser's substantial frequency drift in response to temperature change \cite{Loh2015}. Here, we overcome the challenges of drift by applying a recently developed technique to sense temperature fluctuations of the SBS laser below 100 nK \cite{Loh2019}. We employ this technique to introduce a new feedforward correction scheme, which enables us to compensate for these temperature deviations at the same level of precision. We utilize the stabilized SBS light in the demonstration of a strontium-ion optical clock, which breaks the long-standing clock paradigm of requiring a BCS laser to serve as the master oscillator. Through a clock self-comparison measurement, we achieve a fractional frequency stability of $3.9 \times 10^{-14} / \sqrt{\tau}$ (with the interrogation time $\tau$ in s), beyond the short-term stability achievable by the best microwave clocks. 

For our clock demonstration, we interrogate a {$^{88}$\sr} ion using a fiber-cavity SBS laser with radiation at $1348\unit{nm}$ that is frequency-doubled to reach the narrow-linewidth $^{88}$Sr$^+$ clock transition at $674\unit{nm}$ \cite{Madej2012}. Figure 1a depicts our overall strategy for achieving a stable SBS optical-atomic clock. Starting from two orthogonally-polarized pump lasers input into a high-Q optical-fiber resonator (panel i), the resonator generates two orthogonally-polarized SBS outputs that are Stokes shifted from their respective pumps by ${\sim}12.5\unit{GHz}$ (panel ii). The two SBS lasers interfere on a photodetector to produce a $180\unit{MHz}$ beat note (panel iii), the frequency deviation of which predominantly corresponds to the temperature drift of the SBS resonator \cite{Loh2019}. This method of temperature sensing is based on techniques that connect the differential temperature sensitivity between two orthogonal-polarization resonator modes to a measurement of temperature change \cite{Strekalov2011, Fescenko2012, Weng2014}. However, due to issues associated with both the detection and control of small shifts in temperature, the best prior stabilization efforts have so far been limited to the range of ${\sim}10\unit{\mu K}$ \cite{Lim2019}. Here, as a consequence of the exceptionally narrow SBS lasing linewidth, the resolution of our temperature sensor reaches below $100\unit{nK}$. An even more formidable challenge, however, is the actuation of temperature control with this level of precision, which is complicated by the coupling between temperature and length through thermal expansion. In order to circumvent the need for direct control of temperature, we apply the dual-polarization beat note as a feedforward correction to the SBS laser's frequency (panel iv). This correction brings the SBS laser's stability to a regime where it can be used to interrogate a {$^{88}$\sr} ion optical clock (panel v).


\begin{figure}[t b !]
\includegraphics[width = 0.95 \columnwidth]{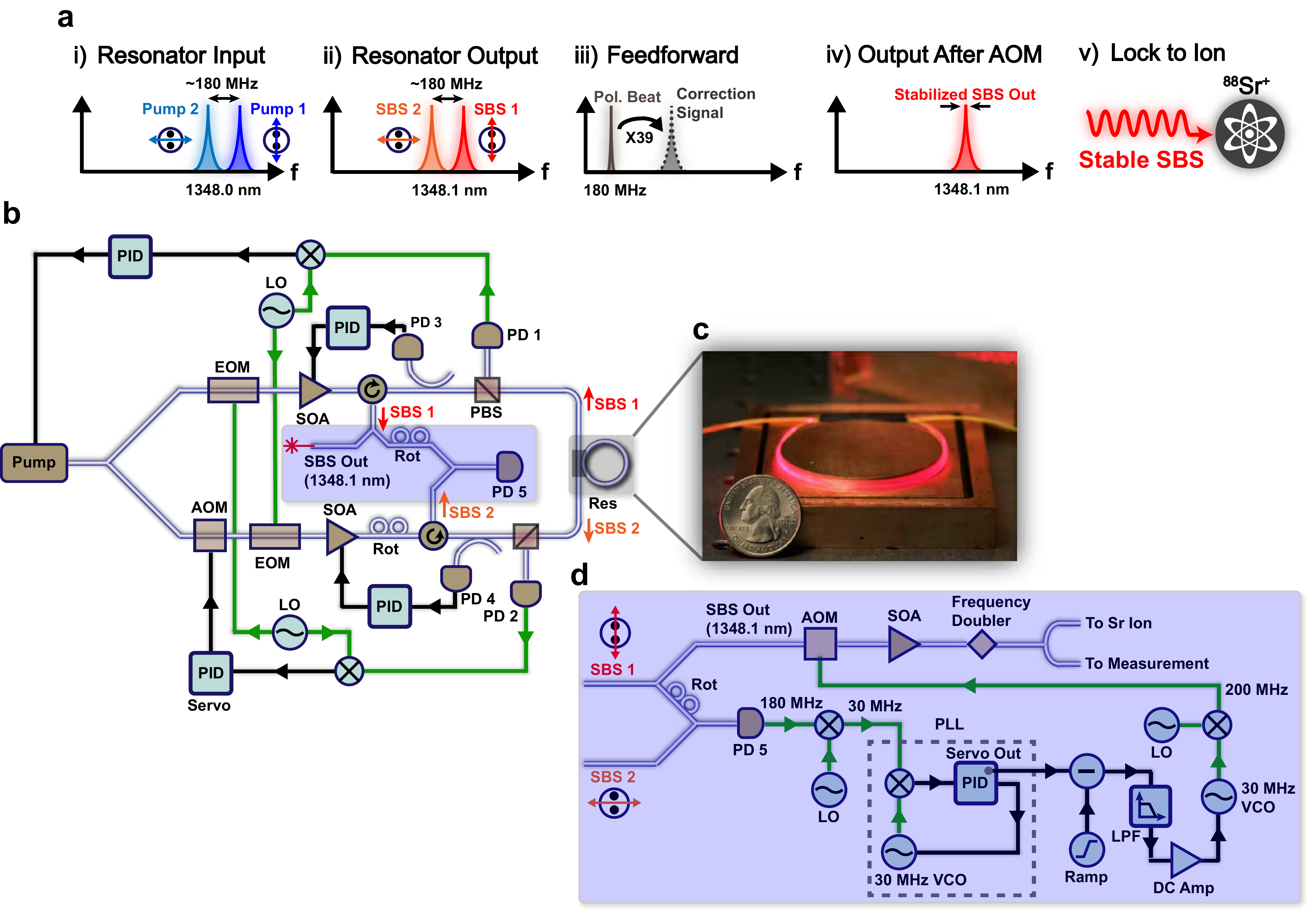}
\caption{
    \textbf{SBS laser setup and stabilization scheme.}
    \textbf{a}, Illustration of the steps required for SBS stabilization. Two orthogonal polarization SBS signals are generated whose beat note is applied to an acousto-optic modulator (AOM) to compensate for the SBS laser's frequency drift.
    \textbf{b}, Diagram of the SBS laser setup. An AOM, electro-optic modulator (EOM), semiconductor optical amplifier (SOA), polarization rotator (Rot), polarization beam splitter (PBS), photodetector (PD), and local oscillator (LO) enable the independent locking of a pump laser to two orthogonal polarization modes of the resonator (Res). Two counter-propagating SBS lasers are created from the pump lasers.
    \textbf{c}, Photograph of the SBS resonator resting within a copper enclosure. The SBS resonator including the enclosure and lid occupies a total space of $3.1\unit{in.}\times3.5\unit{in.}\times1\unit{in.}$
    \textbf{d}, Diagram of the SBS stabilization circuitry. The beat note of two SBS lasers is first converted to a voltage signal through a voltage-controlled oscillator (VCO) operating in a phase-locked loop (PLL). Afterwards, a linear ramp subtraction, low-pass filter (LPF), and factor of 39 multiplication is applied before this signal is used for correcting the SBS laser drift.
}
\label{fig:fig1}
\end{figure}

The SBS laser system (Fig. 1b) consists of a single pump laser at $1348\unit{nm}$ that is split into two separate paths. On one path, an AOM is used both to shift the light by ${\sim}180\unit{MHz}$ and to enable independent control of its frequency. Afterwards, the polarization of one of the pump beams is rotated by $90^{\circ}$, and the two pump beams are then sent into the SBS resonator in opposite directions to accomplish Pound-Drever-Hall (PDH) locking \cite{Drever1983} to the resonator's two orthogonal-polarization modes. The SBS resonator itself exhibits a loaded Q of $1.9 \times 10^{8}$ and consists of 2 m of optical fiber wound around a $2$-inch diameter mandrel that rests within a $3.1 \times 3.5 \times 1$-inch copper enclosure (Fig. 1c). The two pump lasers each generate their own counter-propagating SBS beams, which upon leaving the resonator are both coupled out of the system through a pair of circulators. A portion of the SBS light is also monitored and used to stabilize the amplitude of the SBS laser. Past the circulator, the outputs of both orthogonal polarization SBS lasers are interfered on a photodetector, which serves as our method for sensing temperature change. 

Figure 1d presents the feedforward circuitry used in the stabilization of the SBS laser. After the $180\unit{MHz}$ SBS temperature error signal is generated, a series of operations is applied for the purpose of preparing this signal to serve as a correction to the SBS laser. Using a PLL, the temperature error is first converted to a voltage that is low-pass filtered and amplified before it is converted back again to a RF signal. A net frequency multiplication factor of 39 is achieved through this process, which enables the temperature error to precisely match and cancel the SBS frequency drift. The subtraction of a ramp signal also enables the ability to compensate for a residual linear SBS drift. After correction, the SBS laser output is frequency doubled to reach $674\unit{nm}$ for interrogation of the {$^{88}$\sr} ion.


\begin{figure}[t b !]
\includegraphics[width = 0.95 \columnwidth]{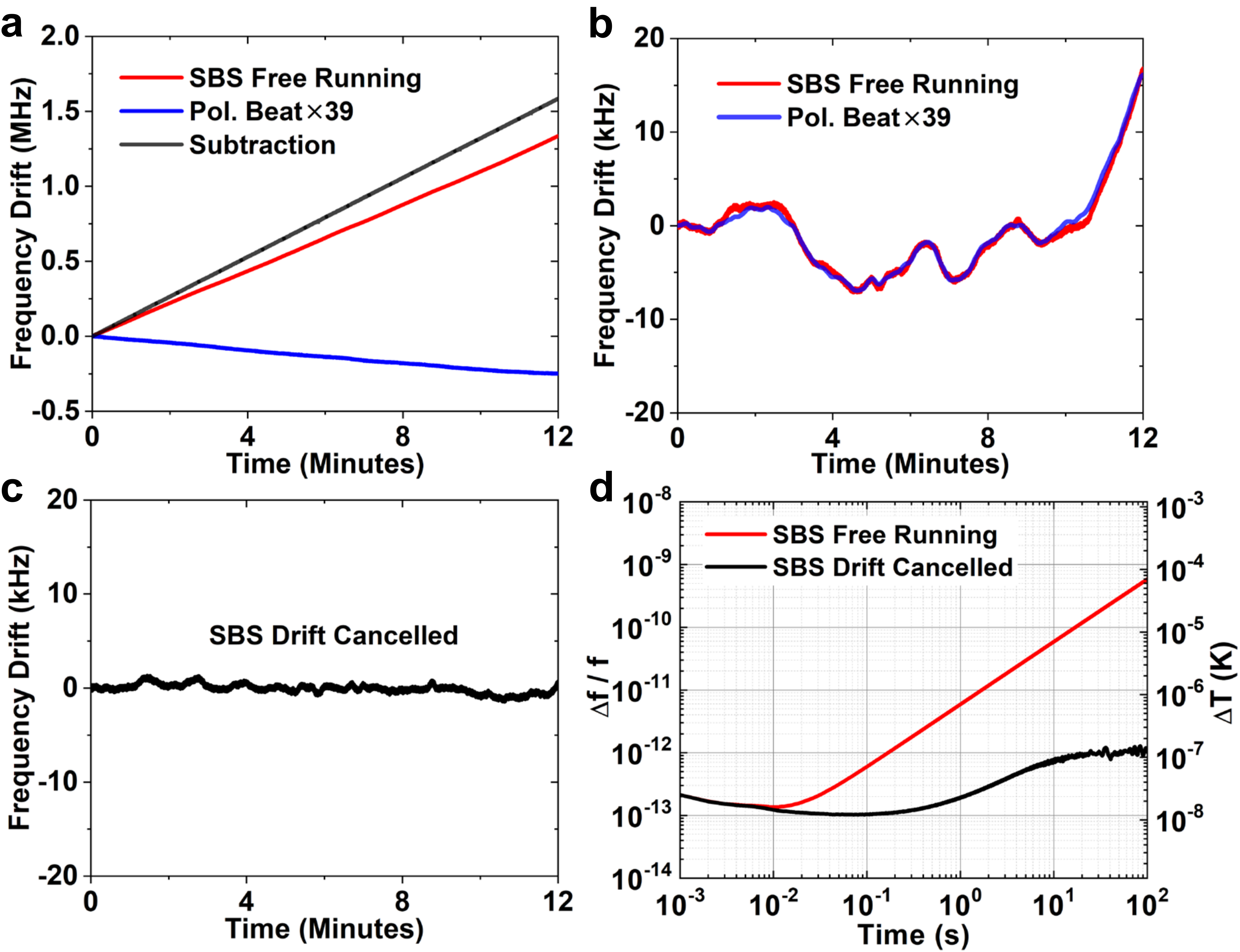}
\caption{
    \textbf{SBS laser drift cancellation procedure.}
    \textbf{a}, Experimental time series of the SBS laser (red) and the polarization beat-note (blue) drift. A numerical subtraction of the two (black) produces a residual linear drift.
    \textbf{b}, Numerical removal of the linear drift from the SBS and polarization beat note time series revealing the underlying temperature-induced frequency drift. The excellent agreement between the two demonstrates that the correction signal accurately tracks the SBS temperature drift.
    \textbf{c}, Subtraction of the SBS and polarization beat-note signals in Fig. 2b. The drift is numerically cancelled through the linear ramp and temperature correction procedure.
    \textbf{d}, Fractional frequency noise measurements of the SBS laser. The resulting noise of the SBS laser with its frequency drift subtracted indicates a long-term frequency variation of $220\unit{Hz}$.
}
\label{fig:fig2}
\end{figure}

Figure 2 demonstrates the feedforward procedure for correcting the SBS laser's drift. Both the frequency of the SBS laser beat note with a reference BCS laser and the frequency of the dual-polarization beat note are measured at time intervals of $1\unit{ms}$ and plotted over a period of 12 minutes (Fig. 2a). The polarization beat is also numerically multiplied by the correction factor of 39 in order to illustrate the correspondence between the correction signal and the SBS laser's frequency drift. Attempting to subtract the correction signal from the SBS laser's drift yields a residual linear drift of $160\unit{kHz}/\unit{minute}$ for the SBS laser frequency. We attribute this linear drift to a slow relaxation of the SBS resonator over time, which is projected to equilibrate on the time scale of months. Upon numerically removing this linear drift from the SBS and polarization-beat traces, the underlying temperature-induced drift of the SBS laser frequency is revealed (Fig. 2b). The correction signal, which now occupies a span of 30 kHz, shows excellent agreement with the remaining SBS laser drift for cancellation. When the polarization-beat frequency is subtracted from the SBS laser frequency, a stabilized SBS laser frequency with $\sim$ 220 Hz frequency fluctuations is obtained (Fig. 2c).

Figure 2d shows the fractional frequency noise derived from the time series traces of Figs. 2a--c. The free-running SBS reaches a minimum noise level of $1.4\times10^{-13}$ (corresponding to a linewidth of $30\unit{Hz}$) at $10\unit{ms}$ but becomes unbounded at longer time scales and increases to $5.8\times10^{-10}$ at $100\unit{s}$. When a numerical feedforward is applied to compensate for both the linear and temperature-induced drift, the SBS laser maintains its performance of $1.0\times10^{-13}$ ($22\unit{Hz}$) at short time scales and experiences a $
>$500-fold reduction in drift over the long-term. The SBS laser frequency excursions become bounded at the value of $1.0\times10^{-12}$ or 220 Hz, which corresponds to temperature stabilization of the SBS laser at a level below $120\unit{nK}$.


\begin{figure}[t b !]
\includegraphics[width = 0.95 \columnwidth]{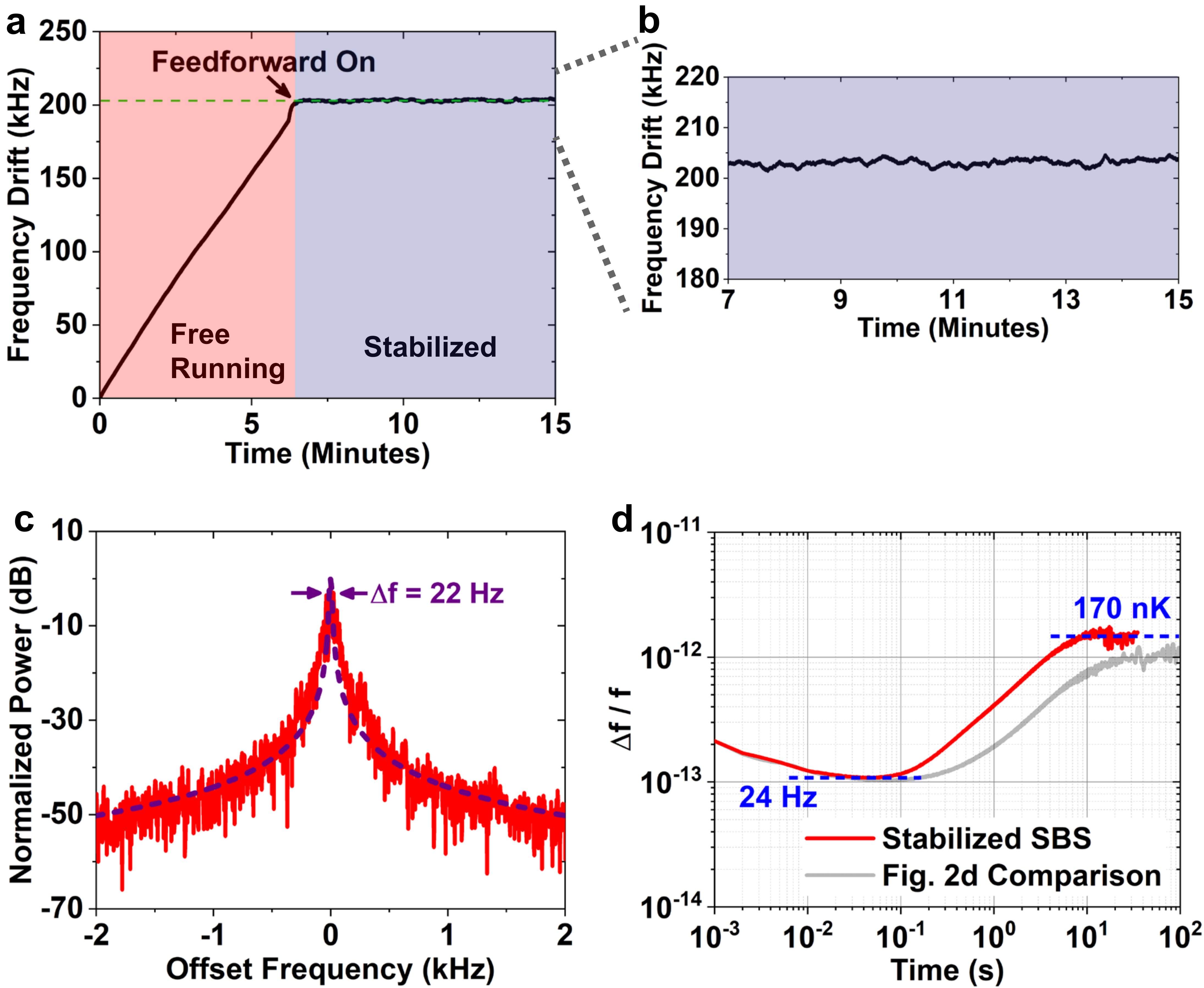}
\caption{
    \textbf{SBS laser stabilization.}
    \textbf{a}, Measurement of the SBS laser's frequency drift (solid line) with and without feedforward stabilization. The measurement interval was $1\unit{ms}$.
    \textbf{b}, Zoomed-in plot of the SBS frequency drift. The average long-term frequency deviation is $310\unit{Hz}$.
    \textbf{c}, Measured lineshape of the stabilized SBS laser (red solid line) and corresponding Voigt fit (gray dashed line). The spectrum is taken with a sweep time of 150 ms and a resolution bandwidth of $10\unit{Hz}$.
    \textbf{d}, Fractional frequency noise comparison between the stabilized SBS laser (red line) and the ideal drift-cancelled SBS laser of Fig. 2d. Long-term temperature stabilization at the level of 170 nK is experimentally achieved.
}
\label{fig:fig3}
\end{figure}

We experimentally implement the numerical feedforward procedure of Fig. 2 and apply the corrections onto the SBS laser's frequency via an AOM (Figure 3a). The SBS laser starts initially in a free-running state for the first 6.5 minutes of measurement and drifts by $200\unit{kHz}$ within this time. Once the feedforward correction is applied, the SBS frequency drift flattens to a value near zero as indicated by the horizontal dashed line guide. A zoomed-in trace of the stabilized SBS frequency drift (Fig. 3b) shows close agreement to Fig. 2c and indicates that our experimental implementation of the feedforward stabilization accurately compensates for both linear and temperature-induced SBS laser drift.

The lineshape of the stabilized SBS laser (Fig. 3c) demonstrates further the laser's exceptional short-term noise. A Voigt fit to the spectrum reveals a linewidth of $22\unit{Hz}$. This value of linewidth is confirmed by the measured fractional frequency noise of the stabilized SBS laser (Fig. 3d), which reaches a minimum of $1.1\times10^{-13}$ at $60\unit{ms}$ and corresponds to a frequency deviation of $24\unit{Hz}$. At long time scales, the noise level becomes $1.4\times10^{-12}$ ($310\unit{Hz}$), which is slightly larger than the $220\unit{Hz}$ excursions found with the numerical drift-cancelling procedure of Fig. 2d. We attribute this difference in drift to noise in the electronics used for feedforward stabilization. At the level of $310\unit{Hz}$, the achieved frequency drift shows our ability to stabilize the SBS laser's temperature fluctuations to within $170\unit{nK}$.


\begin{figure}[t b !]
\includegraphics[width = 0.95 \columnwidth]{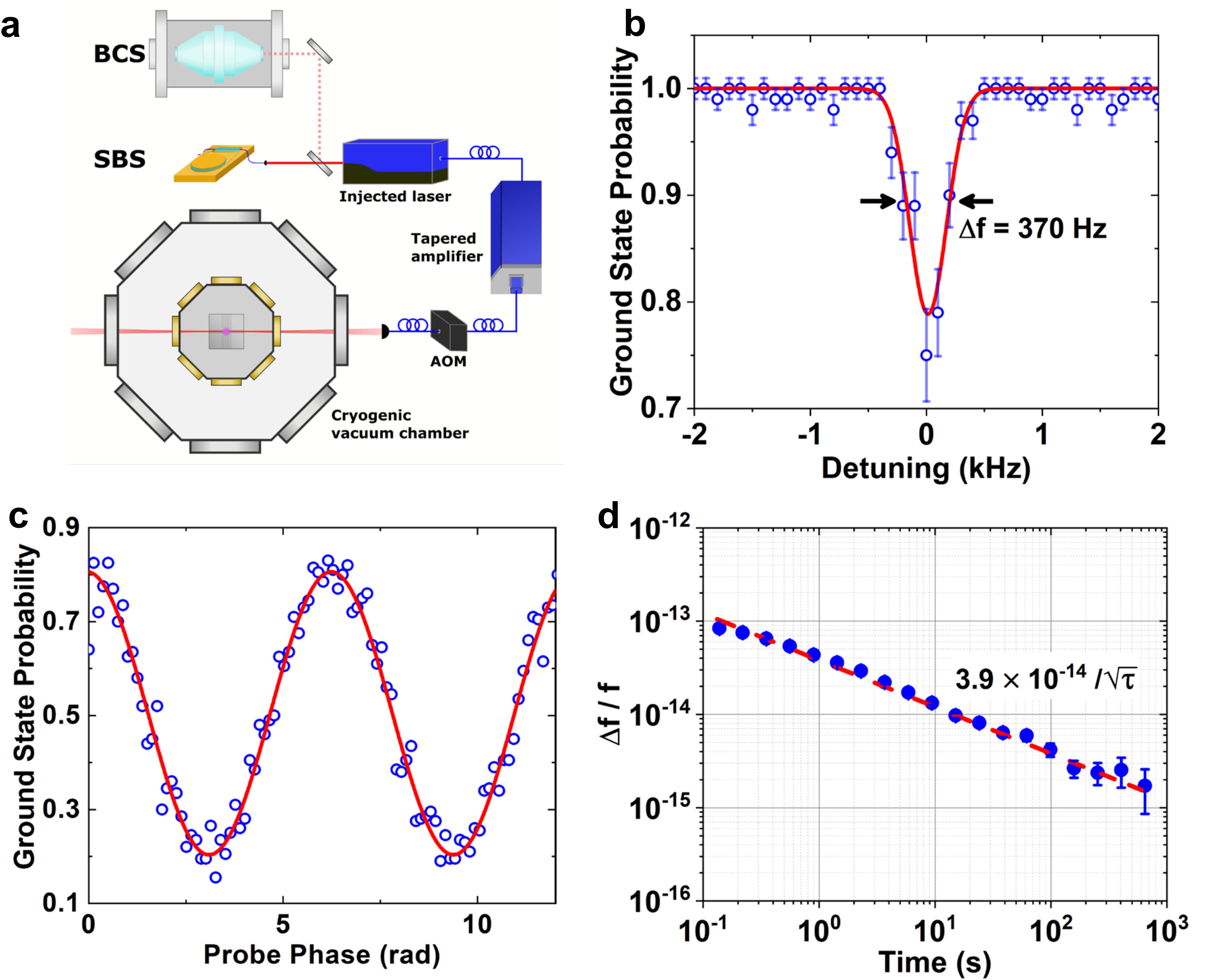}
\caption{
    \textbf{SBS laser optical clock.}
    \textbf{a}, Schematic of laser beam path to the clock chamber.  A flipper mirror allows the SBS laser to be interchanged with a BCS laser as the master frequency source.
    \textbf{b}, Spectroscopy of the $|5S_{1/2}, m_J = -1/2 \rangle \rightarrow |4D_{5/2}, m_J = -3/2 \rangle$ transition in {$^{88}$\sr} measured with the SBS laser, illustrating a coherence-limited linewidth of $370\unit{Hz}$.  The vertical blue bars represent $1\sigma$ error derived from photon counting statistics.
    \textbf{c}, Measured Ramsey fringes on the clock transition.  Two $\frac{\pi}{2}$ pulses are applied around a $\tau=1\unit{ms}$ interrogation period.  The population in the lower clock state is measured as a function of the phase of the second pulse.
    \textbf{d}, Allan deviation of the difference frequency between interleaved clocks.  The measured Allan deviation is divided by $\sqrt{2}$ to estimate the error of a single clock, assuming even distribution of error in the correction signals.  The vertical blue bars indicate the $1\sigma$ error in the Allan deviation \cite{Howe2000}.
}
\label{fig:fig4}
\end{figure}

In order to experimentally demonstrate the practical capability of the SBS laser, we use it to run an atomic clock, stabilizing the laser to the narrow-linewidth $\mathrm{S}_{1/2}\leftrightarrow\mathrm{D}_{5/2}$ quadrupole clock transition in {$^{88}$\sr} ($0.4\unit{Hz}$ natural linewidth). We interrogate a single strontium ion confined $50 \, \mu$m above the surface of a microfabricated surface-electrode trap within a cryogenic ultra-high-vacuum apparatus \cite{Bruzewicz2016}.  To improve the coherence time of the ion's optical transition, we employ a system of passive magnetic field stabilization using persistent superconducting currents \cite{WangCryoTrap2010} and active laser-vibration compensation using an interferometric scheme similar to fiber noise cancellation \cite{Ma1994}.  Using these techniques,  we have measured a coherence time of $2.9\unit{ms}$ with an existing BCS laser.  As shown in Fig. 4a, the clock interrogation light is amplified through a series of injection-locked lasers followed by tapered amplifiers, with fiber-noise cancellation stages to mitigate phase noise picked up as the laser is routed between and across rooms within optical fiber; a single flipper mirror allows us to select either the BCS laser or the SBS laser as the initial seed for the injection stages.

An AOM is used both for scanning the SBS laser over the clock transition and for locking the laser's frequency to the atomic resonance. Spectroscopy of the clock transition performed with the SBS laser reveals a coherence-limited linewidth of $370\unit{Hz}$ (Fig. 4b).  This value of linewidth is consistent with the coherence measured when injecting the optical path with the BCS laser, which we presume to have a narrower linewidth than the SBS laser. Thus we attribute the apparent broadening beyond the $22\unit{Hz}$ capability of the SBS laser to be due to residual uncompensated noise in the optical path, which is independent of the master laser source.

To discipline the SBS laser to the atomic resonance frequency, we create an error signal via a Ramsey experiment consisting of two $\frac{\pi}{2}$ pulses on the clock transition, separated by an interrogation time $\tau=1\unit{ms}$.  As the phase of the second $\frac{\pi}{2}$ pulse is varied, the population in the lower clock state traces out a sine curve (Fig. 4c).  The slope of this signal reaches a maximum when the second pulse is $90^{\circ}$ out of phase with the first pulse. Variations in the laser frequency during the interrogation time result in an additional phase shift and are thus mapped to the ion's state distribution.  Between interrogation cycles, the state of the ion must be detected and then re-prepared into the lower clock level; this leads to a $1.85\unit{ms}$ dead time, during which we are insensitive to frequency fluctuations of the laser.

In order to assess the stability of the Brillouin laser optical atomic clock, we perform a self-comparison measurement via two independently-operated clock signals generated by interleaving distinct sets of correction signals applied to the SBS laser. This technique has been previously used to characterize clock performance when two independent clocks are not available \cite{NicholsonClockComparison2012, NicholsonSystematicClock2015} and is known to accurately capture the short-term clock stability, including ion projection noise and the Dick effect due to dead time in the clock protocol. The self-comparison technique is insensitive to long-term frequency drifts of the ion, which if present would be common to both clock signals, but at the same time underestimates the performance of the clock due to the extra dead time incurred through the interleaving operation. We note that our clock is interrogated on the magnetically-sensitive $|5S_{1/2}, m_J = -1/2 \rangle \rightarrow |4D_{5/2}, m_J = -3/2 \rangle$ transition, which would limit long-term stability if magnetic field drifts were not cancelled; however, a simple protocol for $^{88}$Sr$^+$ clocks has been developed \cite{DubeShiftCancel2005} which eliminates both first-order Zeeman and electric quadrupole shifts of the clock transition and can be implemented in future measurements. 

For a $1\unit{ms}$ interrogation time followed by a $1.85\unit{ms}$ recooling and state preparation period, the effective dead time for each clock is $4.7\unit{ms}$ and results in a measured clock stability of $3.9\times10^{-14}/\sqrt{\tau}$ (Fig. 4d). This value of frequency stability agrees well with numerical simulations we performed of the Brillouin laser optical clock using the measured laser noise, which predict the clock to operate at the level of $4.1\times10^{-14}/\sqrt{\tau}$ for $4.7\unit{ms}$ of dead time. The good correspondence between measurement and simulation suggests that under normal clock operation with $1.85\unit{ms}$ of dead time, the clock stability would reach $2.5\times10^{-14}/\sqrt{\tau}$.


In summary, the advances made here to the SBS laser bring the laser noise into a new regime that approaches the level of performance only currently achievable by BCS lasers. When the SBS laser is used to interrogate an atomic system, the combination offers the potential for creating portable optical atomic clocks with stability surpassing that of state-of-the-art microwave clocks by an order of magnitude or more. For the realization of a truly portable clock, future effort is required in miniaturizing the atomic physics package and in maturing the technology of compact frequency combs \cite{Spencer2018} and SBS lasers. Additional benefits in size and vibration tolerance may be gained by transitioning the fiber SBS laser to an integrated platform \cite{Lee2012, Gundavarapu2019} pending further improvements to the linewidth and stability of chip-based SBS lasers. While not directly explored in this work, other promising architectures for ultralow noise lasers \cite{Liang2015, Zhang2019} may also benefit from the techniques of temperature stabilization developed here.


\section{Methods}

\subsection{SBS Laser Setup}

A 1348 nm external-cavity diode laser outputting 30 mW power serves as the pump of the SBS laser. The pump light is split at a 90:10 ratio between two separate paths that are used to probe the two orthogonal polarization modes of the optical resonator. The 90\% path passes through an additional AOM, which equalizes the power between the two paths. Each path is also phase modulated and amplified such that the power reaches ${\sim}10\unit{mW}$ at the resonator input, which produces ${\sim}5\unit{mW}$ of SBS light at the output. It is essential to the operation of this SBS laser that the counterpropagating SBS light generated by one pump travels alongside the opposite polarization pump that passes through the resonator. The PBS that follows after the resonator then separates the SBS light from that of the orthogonal polarization pump. The pump light is demodulated and used for PDH locking to the cavity resonance. 10\% of the output SBS power is tapped off and used to servo the SBS laser's amplitude, while the remaining 90\% is redirected out by a circulator. The resonator itself sits within a copper enclosure comprising a copper platform that rests on a layer of Viton. The enclosure isolates the resonator from the environment but permits the ability to control the resonator's temperature through the copper.

\subsection{Feedforward Implementation}

In order to stabilize the SBS laser via feedforward, a frequency scaling factor of 39 must be applied to the dual-polarization beat note. This factor is tuned to match the long-term frequency movement of the polarization beat to the drift of the SBS laser. Although multiplication by 39 appears to be optimal, the accuracy of the correction signal does not degrade significantly for scale factors of $39\pm2$. In addition to a frequency scaling, the polarization beat must undergo (1) a ramp subtraction to cancel a linear component of the SBS drift and (2) a 0.03 Hz low-pass filtering step that removes excess noise at short time scales. These steps are accomplished by first phase-locking an RF waveform generator (FM deviation = $30\unit{kHz}$ for $\mathrm{V}_\mathrm{in} = \pm2.5\unit{V}$) to the polarization beat. Owing to the intrinsically lower noise of the waveform generator at microwave frequencies, the servo output becomes a voltage replica of the information contained in the polarization beat note. After subtracting a ramp and low-pass filtering the result, the voltage signal is sent to a tunable amplifier with its amplification factor set to 2. The output of the amplifier is used to control the frequency of a second signal generator operating with FM deviation = $1.075\unit{MHz}$ and $\mathrm{V}_\mathrm{in} = \pm5\unit{V}$. The combination of the amplification factor along with the chosen FM deviation values produces the scale factor of 39 applied to the polarization beat.

\subsection{Measurement of the SBS Laser}

The SBS laser's frequency drift, noise, and spectrum are all measured by interfering the frequency-doubled SBS output with an independent bulk-cavity-stabilized laser operating at $674\unit{nm}$. After photodetection, a $0-200\unit{MHz}$ signal is generated whose noise characteristics are assumed to all be attributed to the SBS laser. In contrast, the $180\unit{MHz}$ dual-polarization beat note is directly produced by separately interfering two orthogonal-polarization SBS signals. For measuring drift, the frequencies of the SBS laser and dual-polarization beat note are tracked across two frequency counters that are simultaneously triggered. Alternatively, for measuring the spectrum, the SBS signal is mixed to $5\unit{MHz}$ and directly detected on a spectrum analyzer.

\subsection{Clock Protocol and Simulation}

Our clock protocol consists of a Ramsey measurement, with initial $\pi/2$ pulse driven by the SBS laser, interrogation time of 1 ms, and final $\pi/2$ pulse $90^{ \circ}$ out of phase with the first pulse. Both pulses are amplitude corrected using a composite pulse sequence \cite{Cummins2003} to mitigate amplitude fluctuations. After completing this Ramsey sequence, we measure the ion state via state-dependent scattering of photons on the cycling $S_{1/2} \rightarrow P_{1/2}$ ion transition. After determining which of the two clock states the ion is in, we update a frequency correction applied to the ion as follows:
\begin{equation}
    \Delta f_{i+1} = \Delta f_i + \frac{1}{2\pi}\frac{\alpha \, m_i}{\tau},
\end{equation}
where $\tau = 1$ ms is the interrogation time, $\alpha \approx 0.2$ is a gain term optimized to achieve the best stability, and $m_i$ represents the measurement result, with $m_i = +1$ ($-1$) corresponding to finding the ion in the $D_{5/2}$ ($S_{1/2}$) state.

In order to implement our interleaved clock self-comparison, we run two independent control loops according to the above protocol. During odd-numbered interrogations, we use and update the first frequency correction $\Delta f^{(1)}_i$; during even-numbered interrogations we use and update the second frequency correction $\Delta f^{(2)}_i$. After the experiment is over the two frequency series $\Delta f^{(1)}$ and $\Delta f^{(2)}$ are compared against one another to compute the Allan deviation of the interleaved clock.

The performance of the interleaved clock is compared against a numerical simulation we developed. This simulation models noise of the SBS laser using measured data for timescales of 1 ms and above (taken from comparisons against a BCS laser) to which is added a white noise term with spectral density of 7 Hz$^2$/Hz to simulate short-time-scale fluctuations of the laser frequency. In the simulation, the laser frequency is updated once per clock cycle based on ion measurements that take into account quantum projection noise and dead time effects. Figure S.1 shows the measured Allan deviation of the unlocked SBS laser, the simulated clock operation in self-comparison mode, and the equivalent Allan deviation that the simulation predicts if a single clock were run with interrogation time of 1 ms and dead time of 1.85 ms. The simulated clock self-comparison finds Allan deviation of $4.1 \times 10^{-14}$ at 1 s, in excellent agreement with our experimental result of $3.9 \times 10^{-14}/\sqrt{\tau}$. The same model predicts that a single clock interrogated by the SBS laser would achieve stability of $2.5 \times 10^{-14}$ in 1 s.


\section{Data availability}

The data sets that support this study are available on reasonable request.


\section{Acknowledgements}

We thank A. Libson, G. V. West, and Prof. I.L. Chuang for helpful discussions.  This work was sponsored by the Under Secretary of Defense for Research and Engineering under Air Force contract number FA8721-05-C-0002. Opinions, interpretations, conclusions, and recommendations are those of the authors and are not necessarily endorsed by the United States Government.

\section{Contributions}

W.L. and J.S. conceived, designed and carried out the experiments with the SBS laser.  W.L., J.S., D.R. and R.M. conceived, designed and carried out the experiments with the clock protocol.  All authors discussed the results and contributed to the manuscript.

\section{Competing interests}

The authors declare no competing financial interests.


\clearpage

\section{Extended data figures and tables}

\subsection{Extended Data Fig. 1 SBS Laser Noise}

\begin{figure}[H]
\centering
\includegraphics[width = 0.5 \columnwidth]{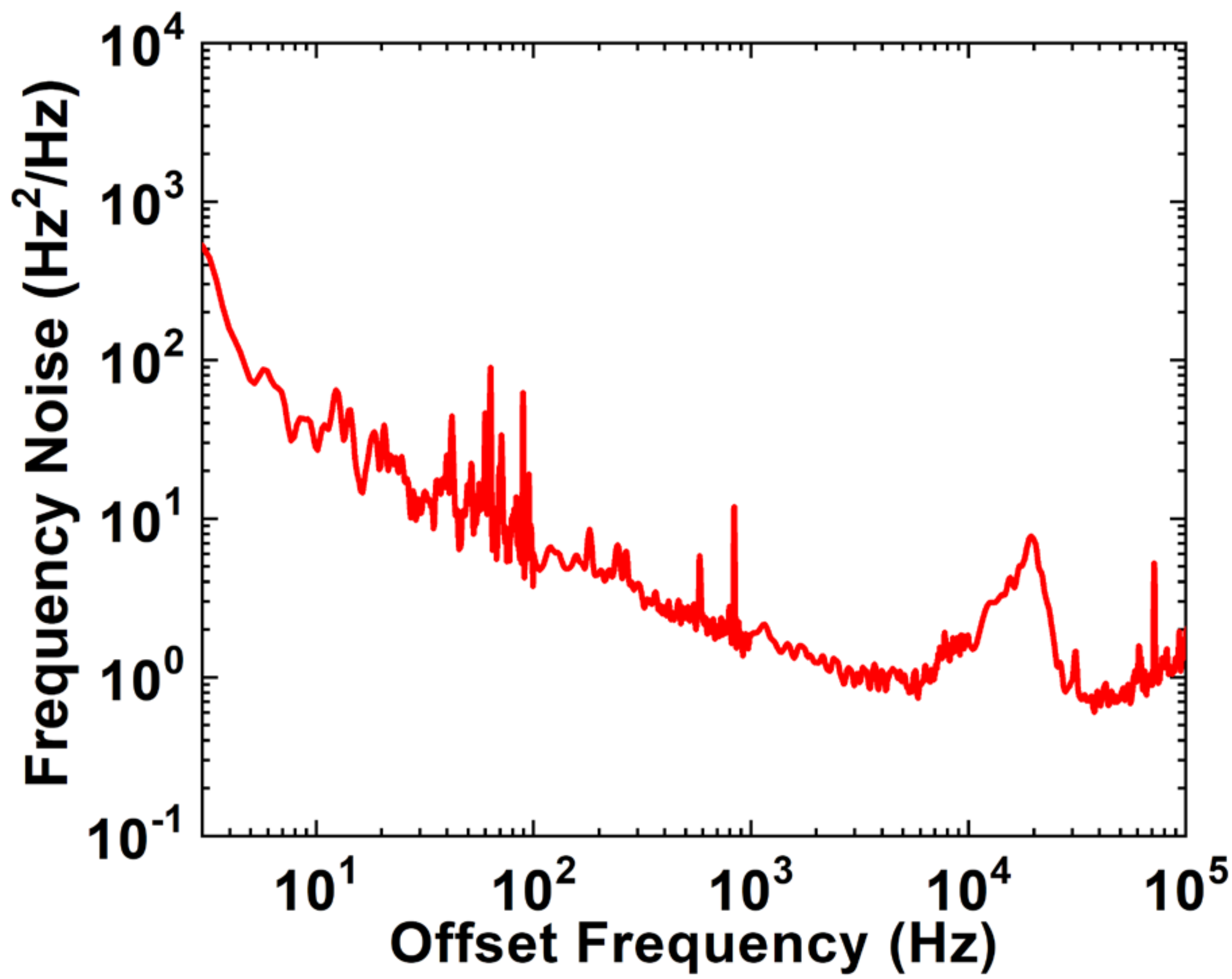}
\caption{
    Measurement of the SBS laser’s frequency noise featuring a white noise floor of $0.7\unit{Hz}^2/\unit{Hz}$ and a gradual increase in noise at lower offset frequencies. An integration over the noise spectral density yields a linewidth of $21\unit{Hz}$.
}
\label{fig:extendeddata_fig1}
\end{figure}

\clearpage

\subsection{Extended Data Fig. 2 Optimization of feedforward stabilization}

\begin{figure}[H]
\centering
\includegraphics[width = 0.95 \columnwidth]{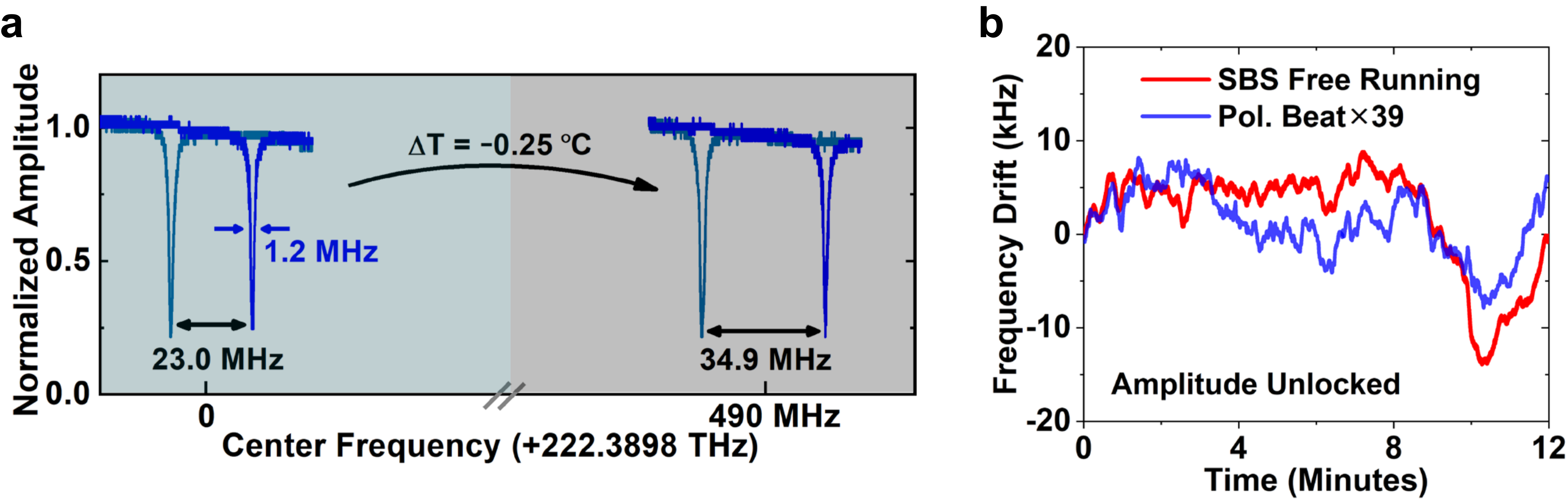}
\caption{
    \textbf{a}, Measurement of the differential temperature sensitivity of the SBS resonator’s two orthogonal-polarization modes. For an applied temperature shift of $-0.25^{\circ}\unit{C}$, the center frequency changes by $490\unit{MHz}$, while the mode separation changes by $11.9\unit{MHz}$. This corresponds to a feedforward correction ratio of 40:1.
    \textbf{b}, Time series trace of the SBS laser’s frequency with the linear drift removed and the SBS amplitude unservoed. The lack of correspondence between the SBS and polarization beat note indicates the inability to cancel frequency drift when amplitude noise is present.
}
\label{fig:methods_fig2}
\end{figure}

\clearpage

\subsection{Extended Data Fig. 3 Numerical simulation of clock performance}

\begin{figure}[H]
\centering
\includegraphics[width = 0.5 \columnwidth]{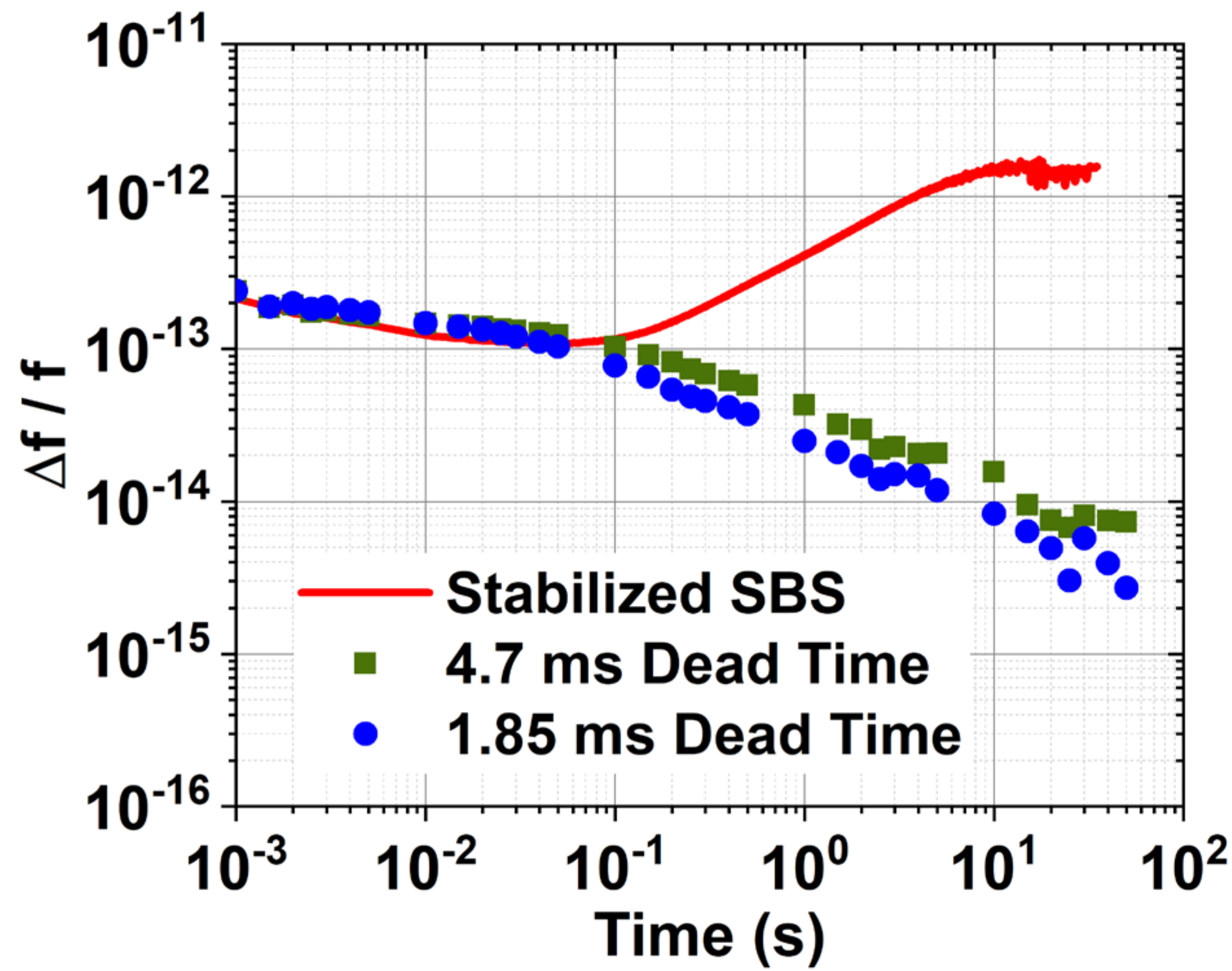}
\caption{
    \textbf{Numerical simulation of clock performance.}
    Measured performance of the stabilized SBS laser (Allan deviation shown as red curve) is used as an input into a clock protocol simulation incorporating projection noise and dead time. The simulation accurately predicts the measured clock performance via the interleaved self-comparison (green squares) and predicts a single-clock stability of $2.5 \times 10^{-14}/\sqrt{\tau}$ (blue circles).
}
\label{fig:methods_fig3}
\end{figure}


\clearpage

\bibliography{SBSClock_biblio_v1}

\end{document}